\documentclass[preprint,12pt]{elsarticle}

\usepackage{amssymb}
\usepackage{pifont}% http://ctan.org/pkg/pifont
    \newcommand{\cmark}{\ding{51}}%
    \newcommand{\xmark}{\ding{55}}%
\usepackage{booktabs}
\usepackage{enumitem}
\usepackage{tabularx}
\usepackage{makecell}
\usepackage{float}
\usepackage{graphicx}

\usepackage{hyperref}

\journal{Computers in Industry}

\begin{document}

\begin{frontmatter}


% \title{Through the buzzword of digital twins: towards a conceptual approach to unify notions, classification and level}
% \title{Behind the buzzword: a maturity level to distinguish among different concepts of digital twins} 
% \title{From buzzword to benchmark: a cross-disciplinary evaluation framework for Digital Twins} 
% \title{Clarifying buzzword for Digital Twin: a novel generic evaluation framework} 
\title{Demystifying Digital Twin buzzword: a novel generic evaluation model}


\author[inst1]{Zhengyu LIU}
\author[inst1]{Sina NAMAKI ARAGHI}
\author[inst1]{Arkopaul SARKAR}
\author[inst1]{Mohamed Hedi KARRAY}

\affiliation[inst1]{organization={Laboratoire Génie de Production de l'Université de Technologie de Tarbes, Université de Toulouse},  addressline={47 Av. d'Azereix}, city={Tarbes},postcode={65000},country={France}}

%$\author[inst2]{Author Two}
%\author[inst1,inst2]{Author Three}

%\affiliation[inst2]{organization={Department Two},%Department and Organization
%            addressline={Address Two}, 
%            city={City Two},
%            postcode={22222}, 
%            state={State Two},
%            country={Country Two}}

\begin{abstract}

Despite the growing popularity of digital twin (DT) developments, there is a lack of common understanding and fundamental definition of DT. This issue creates even more challenges in evaluating the applicability and maturity of existing DT frameworks. A shared understanding of DT based on unambiguous characterization is required to be developed to mitigate the challenge of both research and commercial communications. With these in mind, the objective of our study is to assess the existing DT models from various domains to unify the knowledge and understanding of DT developers and stakeholders. We conducted a systematic literature review and analyzed 25 selected papers out of 107 to identify and discuss the characteristics of existing DTs. 
The review shows an inconsistency in the characterisation of the existing DT models, mostly influenced by case-specific focus of the developers. Therefore, this article proposes a three-phase evaluation framework to assess the maturity of digital twins across different domains, focusing on the characteristics of digital models. The four identified dimensions in this model are \textit{Capability}, \textit{Cooperability}, \textit{Comprehensiveness}, and\textit{Lifecycle}. Based on that, a maturity score is calculated with a implemented weight mechanism to adapt the importance of each dimension for different application requirements. Several case studies are devised to validate the proposed model in general, industrial and scientific cases. 

\end{abstract}

% A RAPID ABSTRACT WRITTEN by one of your nightmares' monster
% Despite the growing interest in digital twins (DT) developments, there is a lack of common understanding and definition for important concepts of DT.  With these in mind, the objective of our paper is twofold. At first, we reviewed the body of knowledge regarding DT developments in manufacturing and extracted as much as possible common grounds for identifying significant aspects of DT. To achieve this, we conducted a systematic literature review and analyzed 25 selected papers to identify and discuss the characteristics of existing DTs. Secondly, this article introduces a four-dimensional evaluation framework to assess the maturity of digital twins across different domains with a focus on the characteristics of digital models. The four identified dimensions in this model are \textit{Capability}, \textit{Cooperability}, \textit{Coverage}, and\textit{ Life-cycle}. Additionally, a weight mechanism is implemented inside the model to adapt the importance of each dimension for different application requirements. Several case studies are devised to validate the proposed model in both scientific and general cases. 
%-----------------

\begin{highlights}
    \item Systematic literature review that reveals DT evaluation inconsistencies.
    \item A generic four-dimensional DT classification schema.
    \item Two fundamental conditions and a quantitative maturity assessment.
    \item Validation through multi-disciplinary case studies.
% \item A systematic Literature analysis reveals inconsistencies and case dependencies in evaluating DT, highlighting the need for a standardized framework.
% \item An evaluation framework based on four key dimensions: Capability, Cooperability, Comprehensiveness, and Lifecycle, built to assess the maturity of DTs across diverse applications.
% \item Two entry conditions and quantitative approach to enhance the proposed evaluation and calculate both dimensional and overall maturity scores.
% \item Case studies to validate the evaluation framework's utility, ensuring its multi-disciplinary applicability.
\end{highlights}


\end{frontmatter}


\section{Introduction}
\label{sec:1intro}

% % For citations use: 
% %       \citet{<label>} ==> Jones et al. [21]
% %       \citep{<label>} ==> [21]
% % Merci
% \begin{figure}[H]
%     \centering
%     \includegraphics[width = 0.8 \linewidth]{Capture.PNG}
%     \caption{Caption}
%     \label{fig:enter-label}
% \end{figure}

\begin{figure}
    \centering
    \includegraphics[width=0.8\linewidth]{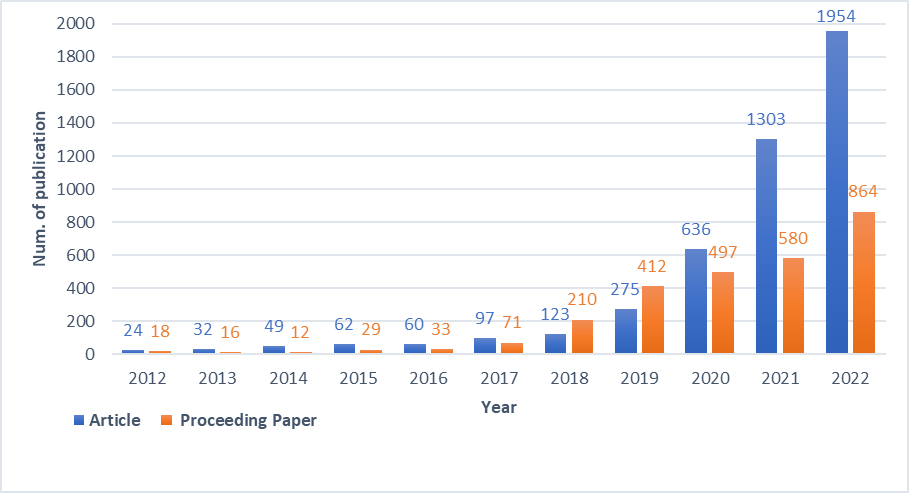}
    \caption{Number of publications of DT from 2012 to July 2023 from the Web of Science, using query \textit{((TS=(digital twin)) OR (QMTS=(``DIGITAL TWINS")) OR (QMTS=(``DIGITAL TWIN DT"))) AND (DT==(``ARTICLE"))}}
    \label{fig:fig1}
\end{figure}

The popularity of \textit{Digital Twins (DT)} continues to grow in the past decade in both academia and industry. Figure \ref{fig:fig1} illustrates the growth in the number of publications since 2012 based on a query on the keyword ``digital twin" on the Web of Science. The corresponding publications saw exponential growth, with under 100 papers yearly until 2018. From 2018 to 2022, over 6000 articles were published, with 1954 articles in 2022, signifying a 20-fold increase in five years. In industry, DT is drawing the attention of the same level. For example, it is the most trending topic in the manufacturing execution system from 2019 to 2021 \cite {shojaeinasab_intelligent_2022}. 

Nonetheless, there is a considerable lack of alignment in comprehending fundamental concepts and definitions of DT \cite{newrzella_5-dimension_2021}. In the young history of DT developments, we have identified two main approaches to identifying the architecture of DT. The first architecture was proposed by Grieves, who published the initial white paper of DT in 2014 \cite{grieves_digital_2014}. Based on the common ground of the digital twin mentioned above, his architecture consists of three dimensions: \textit{physical entity}, \textit{virtual entity}, and bi-directional \textit{connection} between them. The second architecture is from Tao \cite{tao_digital_2017}, one of the most influential papers in the DT domain, adding two additional dimensions: \textit{Digital Twin Data} and \textit{Service}. In this case, the connection is no longer just between two entities but among each one of the other dimensions. As the data lies at the core of his architecture, the connections include \textit{physical-virtual entity connection}, \textit{physical-data connection}, \textit{virtual-data connection}, \textit{service-data connection}, \textit{physical-service connection}, and \textit{virtual-service connection}. It has to be noted that all the links mentioned are designed to be bidirectional.

Considering these and looking at the body of knowledge on digital twin developments, the understanding of digital twins is diverse. Undoubtedly, variances exist throughout the dimensions of DT, such as the frequency of data exchange and the level of automation. Such an abundances may not always be easily delineated by the initial propositions in \cite{grieves_digital_2014}\cite{tao_digital_2017}, due to their broad and high level characterization. Another notion without a consistent understanding is \textit{Digital Twin Prototype} mentioned by Grieves in \cite{grieves_digital_2014} and further interpreted in \cite{grieves_intelligent_2022}. It is described as a digital model that is ``intended" to be further developed into DT with a corresponding physical entity, suggesting it can exist before an actual physical counterpart is developed. Questions are also posed between DT and similar notions, e.g., digital shadow and digital model. Early works before 2018 are often viewed as more aligned with digital shadow or digital model, owing to the limited development of information exchange between physical entity and virtual entity. The connection between the two entities is often unidirectional or even discontinuous, which does not match the definition of DT in today’s point of view \cite{kritzinger_digital_2018}. 

Therefore, despite the DTs' popularity, many hold the opinion that confusion and differentiated understanding are drawing back the development of DT due to its multidisciplinary complexity \cite{newrzella_5-dimension_2021}. Considering this discrepancy issue in understanding DT's concepts and definitions, this article proposes a maturity framework to assess digital twins across diverse fields.
Thus, the fundamental research question we seek to address is presented as follow: 
%----------------
\begin{itemize}
    \item \textbf{How to provide a unified DT evaluation framework to conform to the understanding of practitioners from different disciplines?}
\end{itemize}
%----------------

Our approach in responding to this question is divided into two steps:
\begin{enumerate}
    \item Analyze the literature on digital twins development in manufacturing through a systematic literature review that follows the PRISMA (Preferred Reporting Items for Systematic Reviews and Meta-Analyses) method \cite{page_prisma_2021}.
    \item Introduce an evaluation model that not only assesses existing digital twins effectively but also offers a framework to align the stakeholders' understanding during the DT design phase.
\end{enumerate}

By the end of this research, we would like to propose a DT evaluation framework focusing on the modeling-related features of the digital twin. This focus comes from the recognition that modeling is the key of DT that differentiates it from other related notions \cite{tao_DTCPS_2019}. Based on that, the proposed framework aims to establish a cross-disciplinary consensus and provide insights to assess DT design and existing DT for both academic research and practical implementation of DTs.

This article presents each aspect of our goal in the following. The detailed methodology and specific queries employed in the review are illustrated in Section \ref{sec: Search_protocol}, of which the results are classified and analyzed as one of the two contributions of the paper in Section \ref{sec: class}. Based on this foundation, a three-phase evaluation framework is introduced in Section \ref{sec: Proposal} as the other main contribution. The three phases are illustrated in detail in the subsections: fundamental conditions in Section \ref{sec: 401}, classification schema in Section \ref{sec: 402}, and calculation of maturity score in Section \ref{sec: 403}.
Several case studies are presented and discussed to validate the utility of the proposed framework in Section \ref{sec: Discussion}. As the final part of this paper, conclusions, limitations, and perspectives for future work are given in Section \ref{sec: conclusion}.

% \section{Search protocol}
\section{Systematic literature review methodology}
\label{sec: Search_protocol}
% Having seen the popularity and gap of digital twins in modern academia and industry, it's imperative to delve into the systematic review that ensures a comprehensive exploration of the existing literature and a global understanding of the multifaceted dimensions of DT evaluation.

\begin{figure}
    \centering
    \includegraphics[width=0.75\linewidth]{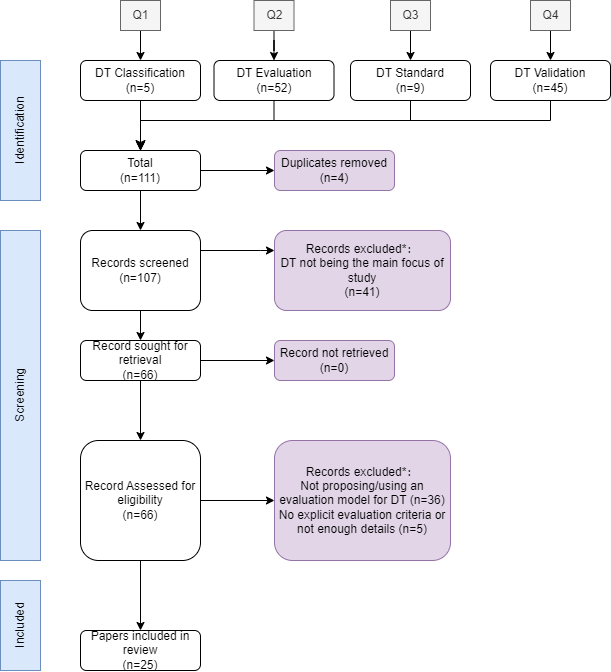}
    \caption{PRISMA workflow showing the identification and screening processes for the literature review}
    \label{fig:f_PRISMA}
\end{figure}

In this section, a systematic review of the literature is conducted to address the gap mentioned in the previous section and to thoroughly understand the current evaluation methodologies. The objective is to look for a comprehensive exploration of the existing research and a global understanding of the involved criteria and dimensions in DT evaluation. The review workflow is shown in Figure \ref{fig:f_PRISMA}, and further interpreted later in this section.

Two research questions are underlined as the main classification criteria of the selected studies, following the main question posed previously in Section \ref{sec:1intro}.
% In order to provide a better answer to the previous questions posed in the introduction, the derivative research questions are specified below: 

\begin{itemize}
    \item What are the existing methods and frameworks used for evaluating digital twin developments, and what are the assessment criteria?
    \item Is there a consensus in the literature for evaluating digital twins generally instead of based on a specific context?
\end{itemize}
% In pursuing an overall understanding of the literature to respond to the research questions, a systematic literature review was conducted guided by the PRISMA method 

Accordingly, the PRISMA (Preferred Reporting Items for Systematic Reviews and Meta-Analyses) method \cite{page_prisma_2021} is identified as the systematic approach for our study in the scope of the Web of Science database. PRISMA is an evidence-based minimum set of items to ensure objectivity and transparency in synthesizing existing literature for systematic reviews and meta-analyses.
The initial search is based on the following keywords, which are considered highly relevant terms and applied in the queries on titles, abstracts, and keywords in the context of DT: classification, evaluation, maturity, standard, and validation. The temporal scope of research is from 2018, which was the start of the upswing growth of DT studies shown in figure \ref{fig:fig1}, to April 2023, when the study of this article began. The queries used, and the numbers of corresponding papers identified are provided in the following list:
\begin{enumerate}[start=1,label={\bfseries(Q\arabic*)}]
    \item for DT classification: ((TS=(``digital twin")) AND TS=(``classification of digital twin" OR ``classify digital twin" OR ``digital twin classification")); the extracted result is\textbf{ 5}.
    \item for DT evaluation: TS=(``twin rate" OR maturity OR ``digital twin level" OR ``level of digital twin")); the extracted result is \textbf{52}.
    \item for DT standard: ((TS=(``digital twin")) AND TS=(``digital twin standard*" OR ``standard* of digital twin" OR ``standard* digital twin" OR ``formaliz* digital twin" OR ``digital twin formalize*")); the extracted result is \textbf{9}.
    \item for DT validation: ((TS=(``digital twin")) AND TS=(``Verification model" OR Compliance OR ``verify digital twin")); the extracted result is \textbf{45.}
\end{enumerate}

An additional justification needs to be added to the query \textit{evaluation}. This research focuses on evaluating the digital twin's maturity instead of evaluating another project using DT as an enabling technology. This is why the word ``maturity" is added to the queries.

Figure \ref{fig:f_PRISMA} shows the workflow with results in each phase. After the initial identification, \textbf{111} papers were found based on the four keywords. By removing the duplicates (i.e., papers extracted repeatedly by different queries), the initial number for screening is reduced to \textbf{107}.

In the screening phase, the exclusion criteria and the number of corresponding excluded papers are listed as follows:
\begin{itemize}
    % \item Papers focusing on a general subject, such as Industry 4.0, where DT is considered a component or enabling technology, are excluded. The number of corresponding exclusions is 41.
    \item Subjects: \textbf{41} papers are excluded in which DT is not the main subject of the article;
    % \item Papers of DT use cases, which lack further comparisons and assessment or without proposing an explicit evaluation strategy, are excluded. The number of corresponding exclusions is 36.
    \item DT use-cases: \textbf{36} papers were excluded because they did not propose any explicit evaluation strategy;
    \item Details: \textbf{5} papers were excluded as they do not provide sufficient details or lack clear arguments for the evaluation strategies.
    % Paper not providing sufficient details or clear arguments directly related to the assessment are excluded. The number of corresponding exclusions is 5.
\end{itemize}
Consequently, \textbf{25} papers have been selected to constitute the main body of research for the literature review.

% ----PRISMA Diag----

\newpage
\section{Classification of selected studies} \label{sec: class}
\begin{figure}
    \centering
    \includegraphics[width=0.5\linewidth]{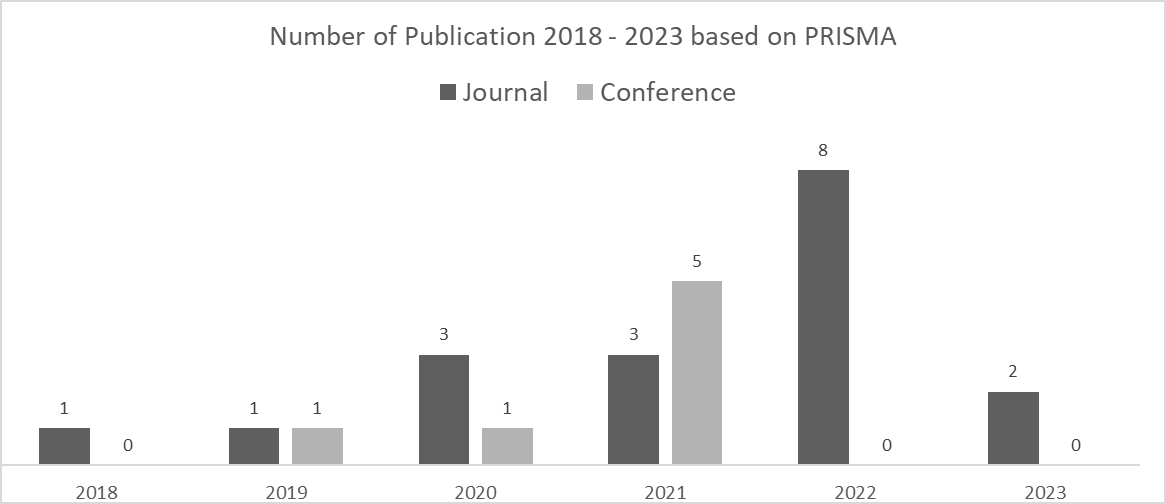}
    \caption{Annual publication of papers on DT evaluation included in the literature review, 2018 - 2023}
    \label{fig:f_numPaperYr}
\end{figure}

This section delved into a detailed analysis of the 25 selected papers. Figure \ref{fig:f_numPaperYr} presents the annual distribution of selected publications within the defined time stamp of our search protocol. As the first step of analysis, it shows accelerating growth in the publication on the DT evaluation, which aligns with the growth curve of general DT articles presented in Section \ref{sec:1intro}.

In most of the existing approaches to evaluate digital twins, critical criteria are identified and selected first, then different methods are applied to assess the maturity of DT based on the chosen criteria. Particularly, these criteria will be named as ``aspects"" in this section, in order to distinguish from the structural dimension in DT architecture and evaluation dimension from the proposal in Section \ref{sec: 402}.

\begin{figure}
    \centering
    \includegraphics[width=0.5\linewidth]{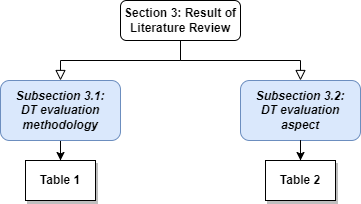}
    \caption{Architecture of Section \ref{sec: class}, the outcome of literature review are presented in two tables following axis \textit{DT evaluation methodology} and axis \textit{DT evaluation aspect}}
    \label{fig:S3architec}
\end{figure}

As shown in Figure \ref{fig:S3architec}, the in-depth analysis is presented by two axis, \textit{DT evaluation methodology} and \textit{DT evaluation aspect} in the next two subsections. \textit{DT evaluation methodology} studied the global features and mechanisms of the evaluation strategies, whose result is presented in Table \ref{tab:table_evaMethod}. \textit{DT evaluation aspect} addresses the selection of specific criteria and aspects by which the DTs are evaluated. The related findings are shown in Table \ref{tab:DTeva}.
% ----S3 Architecture Diag----

\subsection{DT evaluation methodology}

In general, the assessment of DT is conducted via either qualitative or quantitative measures on each defined aspect. For quantitative measures, some research assigns additional weights or importance to each aspect to enhance the adaptability across varied projects. The evaluation results can be either partial, for example a series of scores on each criteria, or holistic, for instance a single score to assess the overall maturity of the entire DT use case.

Works such as \cite{wang_review_2022}, are not featured in Table \ref{tab:table_evaMethod}. This is because these works primarily focus on reviewing previous studies and exploring a range of evaluation models, which does not align with the format and criteria of the table.

\begin{table}
\centering
\caption{Method-centric literature analysis, focusing on the features and mechanisms of \textbf{evaluation strategies}}
\label{tab:table_evaMethod}
\resizebox{\textwidth}{!}{%
\begin{tabular}{cclrrrrrr}
\hline
\textbf{Index} & \textbf{Year} & \multicolumn{1}{c}{\textbf{Desc}} & \textbf{\begin{tabular}[c]{@{}r@{}}Num.\\ Aspect\end{tabular}} & \textbf{\begin{tabular}[c]{@{}r@{}}Num.\\ Rubrics\end{tabular}} & \textbf{Qualitative} & \textbf{Quantitative} & \textbf{Weight} & \textbf{P/O} \\ \hline
\cite{sun_introduction_2022} & 2022 & \begin{tabular}[c]{@{}l@{}}Propose 4 dimensions:\\ fidelity, interoperability,\\ real-time, scalability\end{tabular} & 4 & 0 & \xmark & \xmark & \xmark & Partial \\ \hline
\cite{hosamo_review_2022} & 2022 & \begin{tabular}[c]{@{}l@{}}Review DT in BIM and \\ FM's interoperability\end{tabular} & 1 & 0 & \xmark & \xmark & \xmark & Partial \\ \hline
\cite{atkinson_taming_2022} & 2022 & \begin{tabular}[c]{@{}l@{}}Propose 3 dimensions:evolution,\\ scalability continuity\end{tabular} & 3 & 0 & \xmark & \xmark & \xmark & Partial \\ \hline
\cite{laamarti_isoieee_2020} & 2020 & \begin{tabular}[c]{@{}l@{}}Propose digital twin framework\\ encompassing data collection\\ from personal health devices \\ in different life cycle\end{tabular} & 2 & 0 & \xmark & \xmark & \xmark & Partial \\ \hline
\cite{jacoby_digital_2020} & 2020 & \begin{tabular}[c]{@{}l@{}}Propose consolidation of \\ DT standards and IoT standards\\ as they are overlapping,\\ focus on semantic interoperability\end{tabular} & 1 & 0 & \xmark & \xmark & \xmark & Partial \\ \hline
\cite{oakes_improving_2021} & 2021 & \begin{tabular}[c]{@{}l@{}}Propose a conceptual structure of \\ four main sections \\ based on DT experience report\end{tabular} & 4 & 14 & \xmark & \xmark & \xmark & Partial \\ \hline
\cite{bauer_towards_2019} & 2019 & \begin{tabular}[c]{@{}l@{}}4 DT stages plus 1 stage \\ each described in different views\end{tabular} & 5 & 0 & \cmark & \xmark & \xmark & Partial \\ \hline
\cite{madni_leveraging_2019} & 2019 & \begin{tabular}[c]{@{}l@{}}4 Sequentially proposed \\ DT Levels of sophistication\end{tabular} & 1 & 0 & \cmark & \xmark & \xmark & Overall \\ \hline
\cite{newrzella_5-dimension_2021} & 2021 & \begin{tabular}[c]{@{}l@{}}Propose a 5-dimension model\\ based on literature review\end{tabular} & 5 & 0 & \xmark & \xmark & \xmark & Partial \\ \hline
\cite{davila_delgado_digital_2021} & 2021 & 3 key levels in manufacturing & 1 & 0 & \cmark & \xmark & \xmark & Overall \\ \hline
\cite{stahmann_digital_2021} & 2021 & \begin{tabular}[c]{@{}l@{}}3-dimension maturity model: \\ continuity and consistency,\\ validation, assessment\end{tabular} & 3 & 0 & \cmark & \xmark & \xmark & Partial \\ \hline
\cite{chen_gemini_2021} & 2021 & \begin{tabular}[c]{@{}l@{}}3-dimension maturity model:\\ purpose, trust, function\end{tabular} & 3 & 27 & \xmark & \cmark & \xmark & Partial \\ \hline
\cite{medina_maturity_2021} & 2021 & \begin{tabular}[c]{@{}l@{}}10-dimensional maturity model\\ in Commercial Aerospace Industry\end{tabular} & 10 & 0 & \cmark & \xmark & \xmark & Partial \\ \hline
\cite{malakuti_emerging_2021} & 2021 & \begin{tabular}[c]{@{}l@{}}Maturity model on data silos\\from level 0 to 4\end{tabular} & 1 & 0 & \cmark & \xmark & \xmark & Overall \\ \hline
\cite{duan_development_2020} & 2021 & \begin{tabular}[c]{@{}l@{}}5-level model modeling, \\ interaction, foreknowing,\\ prediction and co-intelligence\end{tabular} & 1 & 3 & \cmark & \xmark & \xmark & Overall \\ \hline
\cite{wei_off-site_2022} & 2022 & \begin{tabular}[c]{@{}l@{}}5-level DT maturity for\\ four parts of DT models\end{tabular} & 4 & 0 & \cmark & \xmark & \xmark & Partial \\ \hline
\cite{kasper_toward_2022}\cite{steindl_generic_2020} & 2022 & \begin{tabular}[c]{@{}l@{}}A low abstraction level platform,\\ based on 5-dimensional DT\end{tabular} & 5 & 0 & \xmark & \xmark & \xmark & Partial \\ \hline
\cite{uhlenkamp_digital_2022} & 2022 & \begin{tabular}[c]{@{}l@{}}31 ranked characteristics\\ in 7 categories\end{tabular} & 7 & 31 & \xmark & \cmark & \cmark & Both \\ \hline
\cite{yu_energy_2022} & 2022 & \begin{tabular}[c]{@{}l@{}}5-dimensional DT: \\ behavior, connection, likeness,\\ physical scale and time scale\end{tabular} & 5 & 0 & \cmark & \xmark & \xmark & Partial \\ \hline
\cite{hu_new_2023} & 2023 & \begin{tabular}[c]{@{}l@{}}3 dimensions\\ (function, value and reliability) \\ consist of 27 rubrics\end{tabular} & 3 & 27 & \cmark & \cmark & \cmark & Both \\ \hline
\cite{agrawal_digital_2023} & 2023 & \begin{tabular}[c]{@{}l@{}}4 Roles of DT \\ and corresponding autonomy levels\end{tabular} & 4 & 0 & \cmark & \xmark & \xmark & Partial \\ \hline
\end{tabular}%
}
\end{table}

The columns in Table \ref{tab:table_evaMethod} are explained as follows:
\begin{itemize}
    \item \textit{Num.Aspect} indicates how many dimensions or aspects by which a DT is evaluated. It is noted as 1 if it comprises a single general DT level;
    \item \textit{Num.Rubrics} is related to the number of more detailed sub-criteria that belong to the dimensions, aspects or, in some cases, categories. It is noted as 0 if there are no rubrics of a lower level;
    \item \textit{Qualitative} checks if a qualitative approach is included in the evaluation;
    \item \textit{Quantitative} checks whether a quantitative approach is included in the evaluation, such as calculation based on numeric scores;
    \item \textit{Weight} signifies if the model contains a weight mechanism to compare and adjust the importance of different aspects and criteria; 
    \item \textit{P/O} implies whether the DT is evaluated based on each aspect or rubric separately, which is noted as \textit{Partial}, or an overall score/level is given to measure its global characteristic, which is noted as \textit{Overall}. Some cases are noted as \textit{Both}, as they contain both dimensional and global evaluation.
\end{itemize}

Table \ref{tab:table_evaMethod} provides an insight into existing preferences in evaluating DT. Some papers, namely \cite{sun_introduction_2022}\cite{hosamo_review_2022}\cite{atkinson_taming_2022}\cite{laamarti_isoieee_2020}\cite{jacoby_digital_2020}\cite{oakes_improving_2021}, focus more on the architectural components without providing an explicit evaluation methodology, despite outlining the vision for the necessary aspects and capabilities of DT in the proposed. Another similarity worth attention for these papers is that they all choose the partial evaluation based on each aspect instead of a general maturity from the global point of view. By tracing the origin of these articles, we find that they are search results based on the query ``DT standard''. Therefore, these two common features are considered to outline the pattern for DT evaluation related to standardization.

Amongst the remaining studies, ten papers employ qualitative measures, whereas three use quantitative methods \cite{chen_gemini_2021}\cite{uhlenkamp_digital_2022}\cite{hu_new_2023}. Additionally, only one paper \cite{hu_new_2023} combines qualitative and quantitative methods. Remarkably, only two articles \cite{uhlenkamp_digital_2022}\cite{hu_new_2023} integrate a weighting mechanism to adjust the importance of each aspect for different use cases, and both of them apply quantitative approaches. They are also the only two papers that provide both partial and overall results for the evaluation outcome. However, from the global point of view, there is a preference for partial evaluation by each aspect over overall assessment from a global point of view. Statistically, fifteen papers choose partial evaluation, and four employ overall assessment.

\subsection{DT evaluation aspect} \label{sec: 302}

Table \ref{tab:DTeva} shows the outcome of the literature review following the \textit{DT evaluation aspect} axis. It summarizes the identified criteria and aspects in the existing DT assessment approach. They are classified into ten aspects in the \textit{Aspect} column and are further interpreted as follows:

\begin{table}[htbp]
\centering
\caption{Summary of reviewed paper's selected \textbf{aspects} in the existing literature}
\label{tab:DTeva}
\begin{tabularx}{\textwidth}{@{}lX@{}}
\toprule
\textbf{Aspect} & \textbf{Index} \\
\midrule
\textit{Function} & \makecell[tl]{\cite{bauer_towards_2019}\cite{madni_leveraging_2019}\cite{newrzella_5-dimension_2021}\cite{davila_delgado_digital_2021}\cite{stahmann_digital_2021}\cite{chen_gemini_2021}\cite{medina_maturity_2021}\cite{duan_development_2020}\cite{wei_off-site_2022}\cite{steindl_generic_2020} 
\\ \cite{uhlenkamp_digital_2022}\cite{yu_energy_2022}\cite{hu_new_2023}\cite{agrawal_digital_2023}\cite{sun_introduction_2022}\cite{hosamo_review_2022}\cite{atkinson_taming_2022}\cite{laamarti_isoieee_2020}\cite{jacoby_digital_2020}\cite{oakes_improving_2021}} \\
\textit{Reliability} & \cite{chen_gemini_2021}\cite{hu_new_2023} \\
\textit{Temporal Scope} & \cite{newrzella_5-dimension_2021}\cite{medina_maturity_2021}\cite{duan_development_2020}\cite{steindl_generic_2020}\cite{yu_energy_2022}\cite{sun_introduction_2022}\cite{atkinson_taming_2022}\cite{laamarti_isoieee_2020}\cite{oakes_improving_2021} \\
\textit{Modelling Scope} & \cite{newrzella_5-dimension_2021}\cite{davila_delgado_digital_2021}\cite{medina_maturity_2021}\cite{duan_development_2020}\cite{wei_off-site_2022}\cite{steindl_generic_2020}\cite{uhlenkamp_digital_2022}\cite{yu_energy_2022} \\
\textit{Coverage Scope} &  \\
\textit{Fidelity}         & \cite{sun_introduction_2022}\cite{oakes_improving_2021} \\
\textit{Scalability}      & \cite{sun_introduction_2022}\cite{atkinson_taming_2022} \\
\textit{Autonomy} & \cite{bauer_towards_2019}\cite{stahmann_digital_2021}\cite{medina_maturity_2021}\cite{malakuti_emerging_2021}\cite{uhlenkamp_digital_2022}\cite{agrawal_digital_2023} \\
\textit{Update Frequency} & \cite{bauer_towards_2019}\cite{davila_delgado_digital_2021}\cite{stahmann_digital_2021}\cite{medina_maturity_2021}\cite{wei_off-site_2022} \\
\textit{Data} & \cite{newrzella_5-dimension_2021}\cite{medina_maturity_2021}\cite{malakuti_emerging_2021}\cite{wei_off-site_2022}\cite{hu_new_2023} \\
\bottomrule
\end{tabularx}
\end{table}

\begin{itemize}
    \item \textbf{Function}: The \textit{function} aspect in digital twin (DT) evaluation refers to the capabilities and services a DT offers. This includes abilities such as information collection, data storage, real-time visualization, synchronization and coordination between the physical and virtual entities, prediction, and decision-making.
    
    Almost all the existing works agree that \textit{function} is a crucial trait to be taken into consideration to assess a DT use case. Some articles used different terms, for example, ``value'' in \cite{hu_new_2023}, which emphasizes application and broader impact, and ``interoperability'' in \cite{sun_introduction_2022} that focuses on ``dynamic interaction'' between the two entities in DT. They are considered included in \textit{function} aspect.
    
    % As almost all the evaluation models name function as a crucial trait, some derive the notion value of DT from the basis of function. The function of DT focuses on DT's capabilites and services, such as information collection, data storage, real-time visualization, and decision-making. The value of DT is considered a dimension separate from function \cite{hu_new_2023}, as it usually evaluates the benefits of DT from a higher level, for example, from the life cycle's view. They are not taken as two independent dimensions in this table, as they both describe DT's capability.
    
    \item \textbf{Reliability}: In the context of digital twin evaluation, \textit{reliability} refers to the quality and dependability of the predictions and decisions made by a DT. This aspect measures how accurately and consistently a DT can simulate, predict, and make decisions.
    
    This criteria  can be vital, especially for DTs of high maturity level, such as intelligent DTs in \cite{madni_leveraging_2019} that utilize supervised or unsupervised learning in complex and uncertain environments. As it focuses more on highly mature DT, \textit{reliability} is not recognized as a widely applied aspect in most evaluation frameworks.
    
    % Since the simulation and prediction are often considered functions and usage of DT, the notion of intelligent DT \cite{madni_leveraging_2019} is proposed, where supervised or unsupervised learning is viewed as the Capability of a highly mature DT. As a result, some define reliability as vital evaluation criteria to evaluate the quality of predictions and decisions made by DT \cite{chen_gemini_2021}\cite{hu_new_2023}. 

    \item \textbf{Temporal scope}: The \textit{temporal scope} aspect, along with other close terms such as life-cycle integration in \cite{medina_maturity_2021}, is another aspect that is frequently applied to evaluate DT. It is related to the life cycle stages where the DT is implemented. The {temporal scope} can range from a single phase, such as the operation of an energy plant, to multiple phases across the entire life cycle. 
    
    This aspect is introduced mainly to address the versatility and adaptability of DTs in different lifecycle stages. This aspect is a key to understanding how DTs are customized to specific applications, especially when they have similar functions and services but not necessarily in the same part of the product lifecycle.
    % Sometimes named as life-cycle coverage, the temporal scope is also listed as an essential trait. It is often related to the phases in the life cycle where the DT is integrated: in some cases, DT focuses on a single phase, e.g., the operation of an energy plant, and in some other cases, DT may cross multiple stages across the entire life cycle. This trait is introduced to better format the DT in application aside from the function trait. DT has been deployed for different purposes in different use cases, with the same or similar functions but not necessarily in the same part of life. 
    
    \item \textbf{Modelling scope}: Aside from the previous aspect, \textit{modelling scope} is another popular trait to evaluate the extent of DT. It examines the architectural hierarchy and potential for aggregation within DT systems. This is a aspect that is designed to consider the various modeling levels at which DTs can operate, from single individual assets to complex, interconnected systems or networks.

    The concept of this aspect originates from Grieves’ proposal of Digital Twin Instance (DTI) and Digital Twin Aggregate (DTA) \cite{grieves_intelligent_2022}. These two notions are introduced to describe the relation between aggregation and individuals of DT in capturing and correlating products' profile in the operation phase. Further work, such as \cite{duan_development_2020}, highlights the importance of identifying hierarchical structures in DT development. This is due to DT's potential to replicate the object in the real-world not just as an isolated entity but as a dynamic part of a larger, potentially interconnected system.
    % Most existing models consider the hierarchy architecture and potential for aggregation of DT, as shown in the modeling scope column in Table \ref{tab:DTeva}, since Grieves’ proposal on DT instance and DT aggregate \cite{grieves_digital_2014}. These refer to a DT's various levels, from a single asset to a complex system of facilities, and also from the vision of interconnected DTs and DT networks. For example, one of the three main pillars from Duan and Tian’s maturity model is based on the hierarchical structure \cite{duan_development_2020}.
    
    \item \textbf{Coverage scope}: The \textit{coverage scope} aspect addresses the informational extent and comprehensiveness of the content covered by a DT. It assesses how thoroughly a virtual entity represents its physical counterpart. 
    
    As stated by \cite{agrawal_digital_2023}, the \textit{coverage scope} of a DT can significantly vary based on its application and the nature of the physical entity. For example, a construction site's DT might focus on geometric and mechanical models, while a DT for power plant would require an additional electric flow model. The coverage scope could be a key indicator of a DT's maturity and applicability. However, it has not been applied as an evaluation aspect in the literature yet.
    % Depending on their specific functionalities and roles, the content of a digital twin can vary considerably \cite{agrawal_digital_2023}. For instance, while geometric and mechanical models may adequately represent a construction site's DT, a power plant's DT necessitates the inclusion of an electric flow model. Given that a DT might not encapsulate all facets of its corresponding physical entity, the "coverage scope" serves as an essential aspect to evaluate the comprehensiveness of a DT.
    
    \item \textbf{Fidelity}: \textit{Fidelity} focuses on the accuracy and likelihood in the mapping process from the physical entity to the digital entity. This aspect assesses how accurately and finely the DT represents specific features of the physical entity. 
    
    They should be explicitly distinguished while related to the previous \textit{coverage scope} aspect. \textit{Fidelity} evaluates the likelihood and accuracy of the virtual replicates' mapping the physical entity's features within the scope defined by the \textit{coverage scope}. This aspect is of extra importance for use cases such as advanced manufacturing and precision engineering. The \textit{fidelity} of DT is considered a crucial trait to evaluate DT's maturity but it is less discussed in the literature.
    
    \item \textbf{Scalability}: This aspect assesses the DT's ability to effectively grow and adapt as the quantity or the scale of the models increases. It focuses especially on the extensibility of integrating the existing virtual models with related entities and formulating an interconnected system. Effective scalability ensures that a DT remains functional and useful even as it expands to cover larger systems, such as a series of vehicles and parallel production stations. \textit{Scalability} could be another aspect worth more attention as it is only applied in very few existing evaluation methods.
    
    \item \textbf{Autonomy}: \textit{Autonomy} refers to the degree of human involvement required in the operation of the DT. It evaluates how independently the DT can function, especially in terms of decision-making, analysis, and execution of actions on a higher maturity level. 
    
    For example, in the framework to evaluate human-machine interaction in DT by \cite{agrawal_digital_2023}, the functions and services are divided into four roles: observer, analyst, decision-maker, and action executor. The level of autonomy is then evaluated based on each role. This aspect would be particularly important for DTs in contexts where rapid and accurate responses are needed, such as in dynamic manufacturing environments or complex system monitoring.
        
    \item \textbf{Update Frequency}: \textit{Update Frequency} is related to the temporal interval of communication and data exchange between the DT's physical and virtual entities. Depending on the use case, this interval can vary widely from milliseconds to days. For example, a production line might require more frequent data exchanges compared to a smart city application. This aspect is also known in the literature as time scale, latency, and frequency. The choice of \textit{update frequency} is a key factor in aligning the DT's capabilities with the practical needs and, therefore a vital aspect to be evaluated in its application.
    % Another aspect concerning time is the update frequency, as the DT relies on exchanging data and information between the physical and virtual entities. The interval between exchanges could scale from milliseconds to days for different use cases. 
    
    \item \textbf{Data}: \textit{Data} is a multifaceted trait that involves data sources, formats, and security in the DT architecture. In Tao's five-dimensional DT architecture\cite{tao_digital_2017}, \textit{data} is listed as a standalone architectural dimension with physical entity, virtual entity, connection, and service. Yet, it receives less attention in evaluation compared to other dimensions in the literature. However, a deeper and more comprehensive evaluation of this \textit{Data} aspect, together with its potential sub-aspects, will be essential because of the evolving landscape of DT application. 
    % Other criteria, such as data and data security, receive less attention. The aspects taken into account include data sources and data formats. In Tao's 5D architecture \cite{tao_digital_2017}, data is listed as an independent dimension along with physical entity, virtual entity, connection, and service. However, in terms of evaluation, it does not have an equal amount of discussion.
    
\end{itemize}
~\
As to conclude from the literature, despite some shared understanding, the existing methodologies and aspects to evaluate DT are diverse and inconsistent as shown in Table \ref{tab:table_evaMethod} and Table \ref{tab:DTeva}. There is no general qualitative or quantitative evaluation across the application domains, possibly due to the diversity of DT, as the requirements and features of DT vary enormously in its widespread use cases. Additionally, most of the existing evaluations are based on a specific context, for example, \cite{chen_gemini_2021} on asset management, \cite{uhlenkamp_digital_2022} on production and logistics \cite{hu_new_2023} on high-end equipment. This universal variation, along with case dependency, underscores the challenge of forming a unified framework for DT assessment.

Another significant obstacle is the absence of a common vocabulary and standardized terminology. Namely, the use of multiple terms for the same concept and the same term used for different concepts across various projects. This confusion would be a serious drawback to the communication in the DT field, and also complicate the cross-disciplinary cooperation to develop highly mature DT.

Furthermore, the previous analysis reveals a notable gap for the \textit{coverage scope}. Evidently, DTs do not always have to map every possible aspect of their real-world objects. This aspect is therefore important to understand the extent of information and how comprehensively the DT mirrors the physical entity. Despite some discussions, \textit{coverage scope} is overlooked in existing evaluation methods without being applied as a standalone aspect. 

The popularity of DT benefits from its diverse, cross-disciplinary application \cite{newrzella_5-dimension_2021}. However, it may not reach its full potential without addressing the previously mentioned gap discovered within the various use cases. A generic and standardized approach is deemed necessary to evaluate DT across the application fields. This is why we propose our three-phase evaluation framework in the coming Section \ref{sec: Proposal}, aiming to provide a consistent and comprehensive basis for designing DTs and assessing existing DTs. 
% As to conclude from the literature, although some common understanding has been reached, the methodology and aspects employed for DT evaluation are still generally diverse. The inclinations and discrepancies are prevalent in the current corpus of literature and highlight the challenge of achieving a unified consensus to assess DT properly. Additionally, the absence of a common vocabulary could cause more complexity for the future development of DT. This is evident in evaluation aspects, where it’s expected that multiple terms are used for the same concept or a single term referring to different concepts in different projects. This lack of standardized terminology can lead to confusion and hinder progress in the field.

% On top of that, DT is not obliged to cover all the aspects of the real-world entity. However, this coverage aspect is not addressed in the literature, as shown in the \textit{coverage scope} column in Table \ref{tab:DTeva}. 

% This is why we are proposing a four-dimensional evaluation framework to evaluate the DT modeling, whose objective is to assess the existing DT on a standard basis and to unify the knowledge and understanding of practitioners before development. 

\section{Proposed framework} \label{sec: Proposal}

% The literature on DT has made significant strides in elucidating many aspects. Yet, one cannot overlook the lack of general evaluation dimensions with common vocabulary and the absence of discussions on the coverage dimension, as highlighted in Table 
% To bridge this gap, we are advancing a four-dimensional evaluation framework in Figure \ref{fig:4dModel_v1}, aiming at simplifying the evaluation process of DT and ensuring that practitioners from different disciplines have a consistent understanding to work together.

Our approach tries to evaluate DT in a general framework from the modeling point of view to establish cross-discipline consensus for evaluation dimensions. The reason for choosing modeling as the central point of view is not only to create a joint base for evaluation. The modeling is deemed the core element of DT, differentiating DT from other related concepts such as Cyber-Physical System \cite{tao_DTCPS_2019}. By prioritizing modeling, our research seeks insights that resonate with the academy and industry in DT implementation.

% The framework combines qualitative and quantitative approaches to provide a distinct evaluation for each dimension. Inspired by multiple-criteria decision-making \cite{taherdoost_multi-criteria_2023}, a weight parameter is assigned to each dimension to accommodate the different importance subject to the actual requirement. In practice, the weight will be scored by the stakeholders and then be normalized for further analysis.

\begin{figure}
    \centering
    \includegraphics[width=0.75\linewidth]{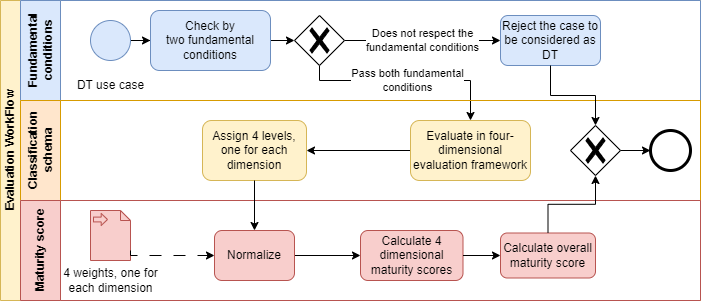}
    \caption{Evaluation workflow in BPMN format including three processes: \textbf{Fundamental conditions}, \textbf{classification schema} and calculation of the \textbf{Maturity score}}
    \label{fig:s4workflow}
\end{figure}

The workflow of the entire proposed framework is presented in Figure \ref{fig:s4workflow} in BPMN (Business Process Modelling Notation) format\footnote{https://www.omg.org/bpmn/}. The entire framework is divided into three phases: \textit{Fundamental conditions}, \textit{Assessment framework} and calculation of the \textit{Maturity score}. Each one of the them corresponds respectively to Section \ref{sec: 401}, \ref{sec: 402}, \ref{sec: 403}. 

% \footnotetext{https://www.omg.org/bpmn/}
As shown in Figure \ref{fig:s4workflow}, the use case must be first examined by the two fundamental conditions to qualify as suitable DT for further evaluation in the next steps. If it fails to fulfill either fundamental condition, then the case is rejected as a proper DT that conforms to the proposed framework.
Once the two conditions are validated, the use case passes to the four-dimensional classification schema as the main qualitative approach to evaluation. A maturity level that best fits the use case's condition will be given on each of the four dimensions. Additionally, four-dimensional weight parameters, inspired by multiple-criteria decision-making \cite{taherdoost_multi-criteria_2023}, have to be assigned to the four dimensions. The objective of weighting is to accommodate further the relative importance of the use case's actual requirement for each specific use case. In practice, the weights will be scored by the stakeholders who develop or intend to develop their DT projects. 
As the final part of the evaluation, the maturity levels and weights will then be normalized to calculate first four dimensional maturity scores, followed by the overall maturity score regarding the entire DT use case. 

\subsection{Fundamental conditions}
\label{sec: 401}
As the starting point of the evaluation, the two fundamental conditions aim at helping the stakeholders, particularly those with limited expertise, in determining the necessity of DT in specific application cases. A DT use case has to validate the two conditions to be evaluated in the next phase. If not, the existing use case is not considered as a DT in the framework. As for DT in the design phase, the stakeholders are suggested to reconsider whether a DT is indeed necessary in their specific application.

These two fundamental conditions originate from a summary of the previous discussions on the difference between DT and other related technologies in Section \ref{sec:1intro}. They are also inspired by the two general architectures proposed by Grieves \cite{grieves_digital_2014} and Tao \cite{tao_digital_2017}. 
While DTs exhibit a wide range of characteristics across its diverse applications, our focus is on identifying the minimal set of characteristics that are universally essential or fundamental to all valid DTs.
A case must first satisfy these two fundamental conditions to be further evaluated within the proposed framework. These two conditions are illustrated in detail as follows:
% and presented later in two preliminary examples for a more thorough explanation.

\begin{itemize}
    \item \textbf{Correspondence}:
    \textit{Correspondence} refers to the alignment between the virtual model and the physical entity.
    This condition is further explored and illustrated through four key elements: 
    \begin{itemize}
        \item \textit{Isomorphism}: In the context of digital twins, isomorphism refers to the structural similarity between the symbolic representations of the physical and virtual entities. Isomorphism is maintained by preserving the structure of relationships, e.g., parthood of components and causality of processes.
        \item \textit{Replicate}: 
        Replicating in DT involves abstracting the physical entity, system, or phenomenon into a preferred digital modeling language. The desired information of a physical entity, system, or phenomenon is captured, reproduced, and represented in the virtual environment by abstraction, forming the digital or virtual counterpart of the physical entity.
        \item \textit{Scope} and \textit{Scale}: 
        Two terms are defined for the informational boundaries on two different aspects represented within the virtual replicate. To distinguish, \textit{scope} is related to the layers of information and the system functionalities, e.g., the plumbing system of a building and the nervous system of the human body. In contrast, \textit{scale} corresponds to the extent of the system represented, ranging from individual components to the entire system.
        \item \textit{Completeness}: 
        Meeting the \textit{correspondence} condition means being complete in the virtual model’s representation of the real entity. \textit{Completeness} demands that the virtual replicate should maintain \textit{isomorphism} to the real entity within the defined \textit{scope} and \textit{scale}. This ensures that the virtual model accurately mirrors all relevant information and structural relationships. By respecting \textit{Completeness} in \textit{correspondence}, the virtual entity captures and maps all the applicable information and structure of the relationship without omission.
    \end{itemize}

    One of the main motivations for choosing this condition for validating DT is the lack of clear causality in the complex system. The cause and consequence might be straightforward in the simple system, whereas such a connection can be veiled in more complex cases. To prevent overlooking any crucial elements, it becomes preferable to ensure \textit{completeness} of \textit{correspondence} during the development of DT.

    \item \textbf{Bidirectional connection}: The bidirectional connection between the physical and virtual entity has been emphasized by two prevailing architectures \cite{grieves_digital_2014}\cite{tao_digital_2017}, which is the distinctive feature between DT and other related technologies. For instances where the primary requirement is monitoring, the connection is unidirectional, and the project is considered a digital shadow rather than a digital twin. Two primary aspects of this condition require further illustration:
    \begin{itemize}
        \item \textit{Autonomy}: Depending on the specific requirements, the bidirectional connection in DT, particulars from the virtual to the physical entity, can range from semi-autonomous to fully autonomous. This variation explains why several academic papers in the previous section incorporate human intervention or automation as critical evaluation dimensions for DT, for example, \cite{medina_maturity_2021}\cite{malakuti_emerging_2021}.
        \item \textit{Continuity}: Rather than employing the term ``real-time'', which depends on specific case requirements and can range from milliseconds to days, we adopt ``continuous" to describe DT's update and interaction frequency. The temporal differences of various use cases are surpassed by describing the update and connection as continuous, and the persisting communication between two entities is highlighted.
    \end{itemize}

    A static digital model, for example, violates \textit{continuity} and will not pass this condition, as it is not updated continuously with the real-world object. Similarly, an isolated simulation model, which operates independently also fails to meet this condition since it simulates without any communication with the real-world object.
    A digital monitoring model matches \textit{continuity} but fails to achieve the \textit{bidirectional connection}. Monitoring only collects information from physical entities without influencing them from the virtual entities. Thus, it is not taken as a DT under this framework. In order to develop a DT, one must integrate continuous and bidirectional exchange and interaction between the two entities.

    % We suppose a motor simulation model in the example to explain \textit{Bidirectional connection}. Starting from the static CAD (Computer-Aided Design) model, which exists without any linkage to its physical counterpart, it is disqualified as a DT. Similarly, a dynamic simulation model, in spite of being active, fails to qualify as it lacks data integration from the actual motor. In the next phase, sensors are installed, and data are collected from the motor and fed to the simulation model so that it mirrors and evolves synchronously with the physical counterpart. With a unidirectional connection, this scenario might be better classified as a digital shadow rather than a digital twin, as it does not satisfy the criteria for a bidirectional connection. Based on the previous steps, if the digital model is capable of influencing the physical entity from the virtual entity, for instance, detecting from the data a symptom of default and triggering deceleration or emergency stop, it is then considered to fulfill the \textit{Bidirectional connection} precondition. Additionally, the operation does not necessitate full automation, and human involvement is not forbidden in this \textit{Bidirectional connection}.
     
\end{itemize}

These two conditions will serve as first-stage guidelines in helping the stakeholders decide if a DT is necessary in their specific application cases. This approach will clarify DT's technical architecture and value proposition and help align strategic goals and operational decisions. It is also useful in avoiding over-expense and redundancy in the early stage of development.

\subsection{Classification schema}
\label{sec: 402}
\begin{figure}
    \centering
    \includegraphics[width=1.0\linewidth]{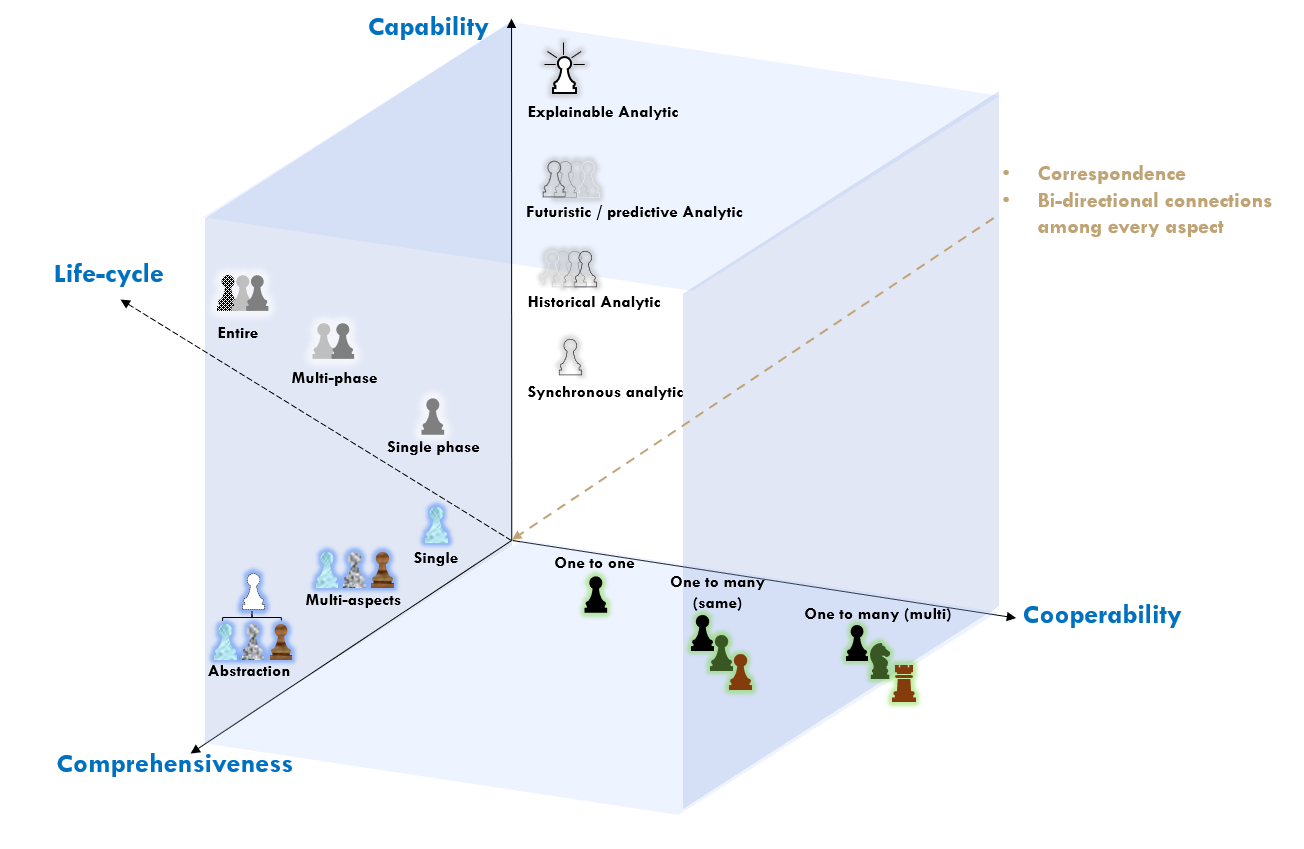}
    \caption{Proposed classification schema, the four dimensions are: Capability, Cooperability, Comprehensiveness, and Lifecycle; the green axis indicates the fundamental conditions in Section \ref{sec: 401}}
    \label{fig:4dModel_v1}
\end{figure}

Once validated by the two fundamental conditions, the use cases will be further assessed in the proposed classification schema shown in Figure \ref{fig:4dModel_v1}.
% To explain the legends used in the figure, the chess icon in the middle represents a physical entity in the real world. The other chess symbols on the four axis signify other virtual models of different features.

The four dimensions in the classification schema come from the summary of the ten evaluation dimensions in Section \ref{sec: 302}.
The selected dimensions focus on the maturity of the fundamental essence of DT, whereas certain dimensions, although valuable for general quality assessment, are not included in the proposed schema. They are often highly case-specific and heavily reliant on the developers' resources.
To be more specific:
\begin{itemize}
    \item Technological sophistication of DT is not considered, which leads to the exclusion of \textit{autonomy} and \textit{update frequency} dimension in the classification schema.
    \item The scale at which DT operates, e.g., an asset or a facility, is not taken as a factor in the evaluation.
    \item The evaluation does not consider DT's precision. Therefore, \textit{fidelity} is not assessed in the framework.
\end{itemize}

Each dimension and corresponding levels are listed and explained in detail in the following:
\begin{itemize}
    \item \textbf{Axis 1: Capability} (Cap): it addresses the application of the DT by providing insights while considering data integration and utility. It originates from the service dimension in Tao's DT architecture \cite{tao_digital_2017} and tries to summarize the \textit{function} and \textit{reliability} dimension in the literature in Section \ref{sec: 302}. \textit{Capability} is identified based on four levels:
    % The \textit{Capability} of DT defined in this framework is distinct from the \textit{Function} dimension of DT in the existing literature. It is a generalization from a broader point of view as the \textit{Function} of DT is considered too diverse to summarize. Initial levels of this dimension primarily focus on data, while advanced levels assess a DT's ability to make explainable decisions. 
    \begin{itemize}
        \item Capability Level 1 (Cap1) -- \textit{Synchronous analytic}: At this level, only real-time data is processed or real-time control is enabled.
        \item Capability Level 2 (Cap2) -- \textit{Historical and descriptive analytic}: At the second level, DT is able to provide analysis based on historical states, such as examining past failures.
        \item Capability Level 3 (Cap3) -- \textit{Futuristic and predictive analytic}: At the third level, DT is able to make predictions of the future state based on technologies like machine learning.
        \item Capability Level 4 (Cap4) -- \textit{Explainable analytic}: At level four, DT can provide the reasoning and logic behind the analysis, prediction, and decision it makes. The process of its operation should be developed to be transparent and comprehensive.
        % \item \textit{Explainable}: DT can explain the reason and logic behind the decision it makes. (Cap5)

        ~\\
        We would like to illustrate this dimension further in the context of a manufacturing plant. At the \textit{synchronous analytic} level (Cap1), DT can monitor and control an assembly line in real-time. At the \textit{historical and descriptive analytic} level (Cap2), the DT now stores and analyzes historical data from the assembly line, such as identifying the failure causes and scheduling bottleneck. DT at the next level, \textit{futuristic and predictive analytic} (Cap3), can predict future defects and schedule maintenance to minimize downtime. As for the\textit{ explainable analytic} level (Cap4), DT will give reasoning processes for the predictive maintenance plan.
    \end{itemize}
    
    \item \textbf{Axis 2: Cooperability} (Cor): It addresses the level of interoperability of the twin with potential digital twin networks. As illustrated in \cite{grieves_intelligent_2022}, the DT can represent either an instance (DTI) or an aggregate (DTA). \textit{Cooperability} is also based on the integration of \textit{modelling scope} and \textit{scalability} dimension in the literature. It has three levels:
    % The reason for establishing this \textit{Cooperability} dimension is rooted in the envisaged interconnection of DTs and DT networks. This vision covers the interactions among diverse assets or devices within a production line to the collaboration of different systems from various domains.
    \begin{itemize}
        \item Cooperability Level 1 (Cor1) -- \textit{One-to-One}: 
        At this initial level, DT is designed to represent a single object. It focuses on modeling a specific product, device, or asset without interaction or extensibility to other digital twins.
        % DT of this level is based on one single object, for example, a product, a device, or a single asset.
        \item Cooperability Level 2 (Cor2) -- \textit{One-to-many in the same environment}: 
        DT represents multiple objects of a certain level of similarity or aggregation. It involves modeling a series of related objects or processes within the same operational environment or context.
        \item Cooperability Level 3 (Cor3) -- \textit{One-to-many in multiple domains}: 
        As for this advanced level, DT is capable of representing  same or same type of objects across different contexts or application domains.
        % DT can model objects from different contexts or applications from multiple domains; for example, DT models the products from the production line via the logistic chain to operation.
        
        ~\\
        As an example, a DT that models a single isolated manufacturing device without links to others is at the \textit{one-to-one} level (Cor1). A DT for a series of production lines in the same manufacturing environment is at \textit{one-to-many in the same environment} level (Cor2). As for another DT tracks products from the manufacturing stage through the logistics chain to the operation is considered at the \textit{one-to-many in multiple domains} level (Cor3), as it covers multiple contexts.
    \end{itemize} 

    \item \textbf{Axis 3: Comprehensiveness} (Com): This dimension means based on which informational scope and coverage we are looking at a physical entity. It describes how many identifiable aspects of the target entity, along with the interactions among them, are covered in the virtual entity. The aspects measures the breadth of information that the virtual entity mirrors from the physical entity.  
    \textit{Comprehensiveness} is also an extension from the \textit{coverage scope} in Table \ref{tab:DTeva}, which is a current gap in the literature.
    % \textit{Coverage} evaluates the comprehensiveness of a DT, namely, the information model to be included and excluded in the DT model, e.g., geometric model, mechanical model, thermodynamics model. It corresponds to the \textit{scope} of the \textit{completeness} precondition while the \textit{scale} is not taken into this dimension as it is considered not closely modeling-related.
    \begin{itemize}
        \item Comprehensiveness Level 1 (Com1) -- \textit{Single aspect}: 
        At this initial level, DT models one single isolated aspect of the object. Assuming a virtual entity mirrors two isolated aspect of the object, it can then be considered two individual DTs of \textit{single aspect}.
        % in the initial level.
        \item Comprehensiveness Level 2 (Com2) -- \textit{Multiple aspects}: 
        At the second level, DT models multiple aspects of the entity. It can integrate and map the interactions between these aspects, allowing a more complex representation and analysis.
        % DT can model multiple connected aspects and combine or switch among them.
        \item Comprehensiveness Level 3 (Com3) -- \textit{Abstraction}: DT operates at a higher aggregation level. It has a holistic view of the entity and is able to look into particular aspects as required. Furthermore, the influence and interaction of each aspect can be analyzed at the abstraction level. This level aligns with the Digital Twin System Interoperability Framework in \cite{budiardjoDigitalTwinSystemInteroperabilityFramework12072021Pdf}. It suggests that DT's holistic and detailed views are a result of its integration into a system of systems, offering both aggregated and specific insights

        ~\\
        For instance, the digital twin of a whole human body is considered at the \textit{abstraction} (Com3) level of the physical entity. However, the digital twin of the cardiovascular system is at the \textit{multiple aspects} (Com2) level as it includes the circulatory system and heart system with the interaction in between. Additionally, the DT of the skeletal system alone is at the \textit{single aspect} (Com1) level.
        % The more detailed we get, we will reach the single aspect.
    \end{itemize}
    \item \textbf{Axis 4: Lifecycle} (Lc): It defines to which extent the DT addresses different phases of an entity's lifecycle. An entity life cycle phase includes the Beginning of Life (BOL), the Middle of Life (MOL), and the End of Life (EOL) \cite{kiritsisResearchIssuesProduct2003}.
    % In \textit{Life-cycle} dimension, DT’s temporal scope is defined according to the product life cycle, from design, prototype, and test to deployment, monitoring, and maintenance. In order to unify the various phases in different contexts, they are classified into beginning of life(BOL), middle of life(MOL) and end of life(EOL) . The other temporal characteristic, update frequency, is not included in the proposal because it is considered more of a case-specific feature than a criterion for evaluating DT.
    \begin{itemize}
        \item Lifecycle Level 1 (Lc1) -- \textit{Single phase}: DT covers one single phase from BOL, MOL and EOL and cannot be directly reused in another phase.
        \item Lifecycle Level 2 (Lc2) -- \textit{Multiple phases}: DT incorporates more than one phase from BOL, MOL, and EOL, and can be reused in another phase.
        \item Lifecycle Level 3 (Lc3) -- \textit{Entire lifecycle}: DT covers all three phases: BOL, MOL and EOL, while the model can switch among them.
    \end{itemize}

    ~\\
    To give an example, a DT of production operation is on the \textit{single phase} level (Lc1), while a DT covers both design and operation is on the \textit{multiple phases} level (Lc2). DT which operates throughout the whole lifecycle of the product, from design to disposal, is at the \textit{entire lifecycle} level (Lc3).
\end{itemize}

However,  the \textit{Comprehensiveness} dimension in the proposed framework has to be distinguished from the notion of \textit{completeness} in the \textit{correspondence} condition in Section \ref{sec: 401}. \textit{Completeness} emphasizes the proper inclusion and representation of all the details in the given \textit{scope} and \textit{scale}. \textit{Comprehensiveness} concerns defining the boundary between inclusion and exclusion of the DT application, which goes with setting up the \textit{scope}.

By the end of this phase, the classification process of DT is finished, and the levels on the four dimensions pass on to the calculation of the maturity score in the next phase.

\subsection{Maturity score}
\label{sec: 403}
As the starting point to obtain the maturity score, the normalized dimensional maturity is calculated based on the DT's corresponding level in this dimension and the total number of levels in this dimension: 
\begin{equation}
    m_i = \frac{n_i}{N_i}
\end{equation}
where $m_i$ is DT's normalized maturity score in dimension $i$, $n_i$ DT's maturity level in dimension $i$, $N_i$ is total number of levels in dimension $i$. $i \in \{Cap, Cor, Com, Lc\}$, which are the abbreviations respectively correspond to dimension \textit{Capability}, \textit{Cooperability}, \textit{Comprehensiveness}, \textit{Lifecycle} presented in Section \ref{sec: 402}. 

\begin{table}
\centering
\caption{The corresponding weight score for difference importance}
\label{tab:weightImp}
\begin{tabular}{c|c}
\hline
\textbf{Importance} & \textbf{Weight Score} \\ \hline
Low & 1 \\
Medium-Low & 2 \\
Medium & 3 \\
Medium-High & 4 \\
High & 5 \\ \hline
\end{tabular}
\end{table}

For example, a DT able to make predictions for a future state is at \textit{Futuristic} level (Cap3) in the \textit{Capability} dimension (4 levels in total), so its maturity score is $0.75$.

The corresponding weight score for each dimension of different importance is shown in Table \ref{tab:weightImp}, scoring from 1 to 5 for low to high importance. The normalized weight of a specific dimension is calculated based on the dimension's weight score and the sum of the total weight score:
\begin{equation}
    W_i = \frac{w_i}{\sum w_i}
\end{equation}
where $W$ is the normalized weight, and $w$ is the weight score determined by stakeholder. $i \in \{Cap, Cor, Com, Lc\}$, respectively correspond to dimension \textit{Capability}, \textit{Cooperability}, \textit{Comprehensiveness}, \textit{Lifecycle}. 

A cap of 0.5 is set as the maximum normalized weight for any dimension to avoid dominance by a single dimension. If a particular dimension reaches the cap, the remaining weights are proportionally distributed among the other dimensions, with their combined total equal to 0.5. For example, for a DT whose weight scores are \{5,2,1,1\}, instead of \{0.56,0.22,0.11,0.11\} the corresponding normalized weights are \{0.5,0.25,0.125,0.125\}.

The objective is to present the level of maturity along with the weight to check the consistency between importance and level of development in each dimension. The vital dimension with a low maturity level would become the focus of future development.

\begin{equation}
    L_{DT} = \sum m_i W_i
\end{equation}

An overall score $L_{DT}$ of maturity for the DT use case can also be calculated from the sum of the multiplication of normalized maturity score $m_i$ and normalized weight $W_i$ of each dimension. They are computed in previous processes for dimensional maturity scores. $i \in \{Cap, Cor, Com, Lc\}$, which signifies the abbreviation of the dimension.

\section{Discussion} \label{sec: Discussion}
In this section, the proposed framework and calculation methodology are tested and validated on various cases, namely industrial cases, general cases, and scientific cases.

\subsection{Industrial case studies}
Both projects in this subsection, Living Heart and Emma Twin, are projects from \textit{Dassault System}, and are categorized as ``Virtual Twin" by the company. ``Virtual Twin" is presented by \textit{Dassault System} as an extension and improvement of DT \cite{netvibesdassaultsystemes02DigitalTwin2023}. The suggested distinction is that while the DT addresses the past and real-time events, the virtual twin can ``predict the future and simulate what-if scenario". This assertion is diverged from many researchers' views in the literature, such as \cite{madni_leveraging_2019} and \cite{tao_digital_2018}, where they claimed that making predictions is within the scope of the capability of DT or is one of its primary services. We think it is essential to validate these two projects under the proposed framework, either being a digital twin or a ``virtual twin''.

\subsubsection{Living Heart}
The Living Heart project from \textit{Dassault System} integrates advanced technologies in biomedical research to study the integrative electro-mechanical response of the whole heart \cite{baillargeon_LivingHeartProject2014}. The process starts with acquiring high-definition images from diverse cardiac diagnostic tools. These images are then transformed into three-dimensional, finite element models to mirror the heart structure. Algorithms and multi-physics simulations are used to replicate the heart's electrical activities, muscular movements, and blood circulation dynamics. Its objective ranges from individualized and customized medical interventions to test new applications and devices to advanced cardiac research.

\begin{figure}
    \centering
    \includegraphics[width=0.5\linewidth]{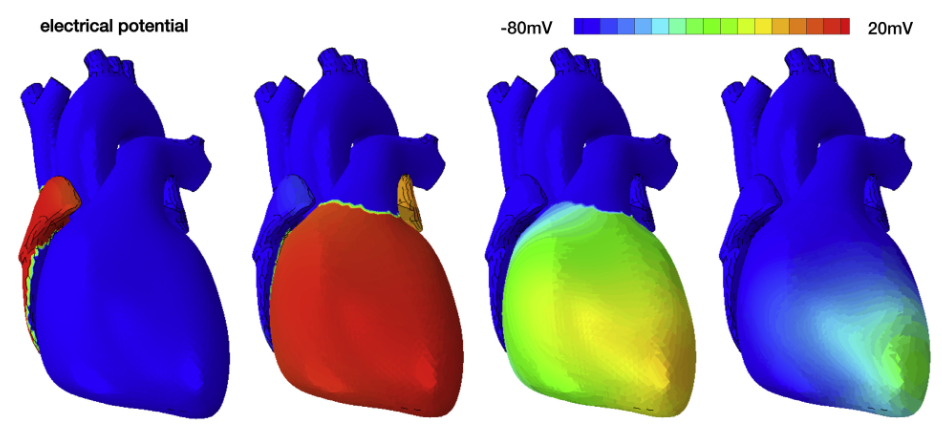}
    \caption{Figures spatiotemporal evolution of mechanical displacement across the human heart from \cite{baillargeon_LivingHeartProject2014}}
    \label{fig:enter-label}
\end{figure}

In this project, the finite model is taken as a replicate of the real heart containing various layers of information, including electrical activities and muscular movements. In this sense, it fulfills the \textit{correspondence} condition in its application scope. However, no clear evidence shows that the virtual replica is updated with real-time data, let alone the Virtual-Physical connection, such as the optimization information. The work in this project focuses more on the simulation and tests without direct and continuous interaction with the physical entity. Therefore, the Living Heart project does not respect the \textit{bidirectional connection} condition. In this case, the project is closer to a digital model than a digital twin.

The mapping between dimensions in DT architecture and Living Heart elements can be described as follows:
\begin{itemize}
    \item \textbf{Physical Entity}: Human heart controlled by the interplay of electrical and mechanical fields;
    \item \textbf{Virtual Entity}: Finite element models to mirror the heart, including all corresponding data in the scope;
    \item \textbf{Bidirectional connection}: No clear continuous connections found in between.
    % \item \textbf{Service}:
    % \item \textbf{data}:
\end{itemize}

As stated previously, the Living Heart project does not fit the fundamental condition for DT in the proposed framework. However, it has the potential to become a practical DT use case with further development to establish and enhance the connection between the physical and virtual entities.

\subsubsection{Emma Twin}
Emma Twin is another project from \textit{Dassault System}, which is an ``accurate virtual representation of a real person" where medical treatments and operations can be tested virtually \cite{MeetEmmaTwin2023}. Emma Twin is said to be created from an anonymous woman’s medical data and characteristics. It can predict and visualize the health condition of multiple parts in its 3D-modelled body, including the Living Heart, which was discussed previously. It can simulate diverse scenarios, such as a heart attack in the middle of a meeting and the outcome of surgery. 

In the scope of an entire human body, we fail to find strong clues suggesting that Emma Twin covers all the concerning parts and elements. We can assume it fulfills the \textit{correspondence} condition to advance the analysis. However, as it is created from a specific person's medical data and tested for various conditions without further monitoring, it could be deduced that there is no constant connection between the physical and virtual entities. 

So, the mapping to the dimensions in DT architecture is incomplete for Emma Twin and does not respect the \textit{correspondence} fundamental condition:
\begin{itemize}
    \item \textbf{Physical Entity}: A real person with the corresponding medical data and characteristics;
    \item \textbf{Virtual Entity}: 3D modeled body in Emma Twin where the tests are conducted;
    % However, the virtual twin is not a complete replicate of the physical entity
    \item \textbf{Bidirectional connection}: No connections once the virtual entity is established.
    % \item \textbf{Service}:
    % \item \textbf{data}:
\end{itemize}

In the case of Emma Twin, it does not fulfill the \textit{Bidirectional connection} condition, and it is not validated as a digital twin in the proposed framework.

\subsection{Application case studies}
In this subsection, two general cases that can be seen in daily life are further evaluated using the proposed framework. Their scoring and weighting are based on a questionnaire during the JNJN (National Day of Digital Twin) event in 2023 \footnote{\url{https://jumeaunumérique.fr}}.

\subsubsection{Google Map Navigation}
As one of the most popular mobile applications being used daily, Google Maps was considered akin to a case of DT mainly because of its navigation functionality \cite{Google02}. This assertion may not be accepted by all, but some aspects of Google Map Navigation align with the concept of DT.  
In general, Google Maps collects data such as vehicle speed and position and the route conditions from the real world to update the map and the navigation plan. The data collected is considered to fulfill \textit{correspondence} condition in the navigation scope to fulfill its guiding purpose. This process can be described as ``a real-time virtual representation of the real world" with continuously updating information, which is the connection from physical entity to virtual entity. 
Optimized navigation plans for better driving are provided based on synchronized information. The driver would decide whether to follow the guideline, while the application will adjust its routing plan accordingly. This could be viewed as a semi-autonomous connection from the virtual entity to the physical entity. Additionally, in an ideally fully automated car, this interaction between virtual and physical would be fully autonomous.

% \subsection{\textbf{ADJUST text figure FOR googleMaps and textla car}}

\begin{figure}
    \centering
    \includegraphics[width=0.5\linewidth]{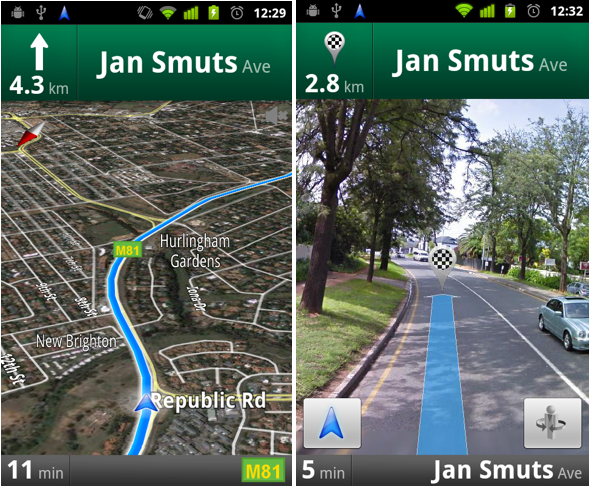}
    \caption{A example of Google map navigation in \cite{Google02} where the guiding is shown in the representation of the real route}
    \label{fig:googlemap}
\end{figure}

In conclusion, the mapping between dimensions in DT architecture and Google Map Navigation elements can be described as follows:
\begin{itemize}
    \item \textbf{Physical Entity}: Routes and vehicles in the real world (noted as PE in the following text);
    \item \textbf{Virtual Entity}: Virtual representation of the routes and cars in the application (noted as VE in the following text);
    \item \textbf{Bidirectional connection}: Real-time vehicle and routing information from PE to VE; optimized traveling plan and guided directions from VE to PE.
    % \item \textbf{Service}:
    % \item \textbf{data}:
\end{itemize}

\subsubsection{Tesla Vehicles}
The American automotive manufacturer \textit{Tesla} claims to utilize DT technology in every produced car. Data collected from sensors in the vehicle allow the company's algorithm to predict potential malfunctions and reduce the maintenance frequency. It also stated that real-time mechanical issues in Tesla Motors can be fixed by downloading over-the-air software updates. Broader DT applications include driver safety and fuel efficiency \cite{piromalisDigitalTwinsAutomotive2022}\cite{eichberger_TeslaDT_2020}.

Based on the previous statement, it is unclear whether Tesla's DT covers the scope of the entire car. However, it can be confirmed that it covers all the corresponding information in the scope of engine and mechanical systems. Therefore, it is considered ``complete" in the scope of DT on the engine and mechanical systems and, therefore, fits the \textit{correspondence} condition. The physical-virtual connection is realized by data collected from sensors, and the virtual-physical connection is based on a predictive algorithm and problem-solving by software updates. It is, therefore, validated as a DT in the proposed framework.

In conclusion, the mapping between dimensions in DT architecture and Tesla vehicle elements can be described as follows:
\begin{itemize}
    \item \textbf{Physical Entity}: Vehicle's engine and mechanical systems (noted as PE in the following text);
    \item \textbf{Virtual Entity}: Virtual representation car including all the relevant details in both systems (noted as VE in the following text);
    \item \textbf{Bidirectional connection}: Data collected from sensors installed on cars; Predictions and downloading solutions to mechanical issues.
    % \item \textbf{Service}:
    % \item \textbf{data}:
\end{itemize}

\begin{table}
\centering
\caption{Maturity calculation for two general cases: Google Map and Tesla Vehicle, $Cap$-Capability, $Cor$-Cooperability, $Com$-Comprehensiveness, $Lc$-Lifecycle, $L_{DT}$-Overall maturity score}
\label{tab:Caseweight02}
\resizebox{\textwidth}{!}{%
\begin{tabular}{l|cccc|cccc}
\hline
\textbf{} & \multicolumn{4}{c|}{\textbf{Google Map}} & \multicolumn{4}{c}{\textbf{Tesla vehicle}} \\
 & \multicolumn{1}{l}{Maturity Level} & \multicolumn{1}{l}{Maturity Score} & \multicolumn{1}{l}{Weight Score} & \multicolumn{1}{l|}{Normalized Weight} & \multicolumn{1}{l}{Maturity Level} & \multicolumn{1}{l}{Maturity Score} & \multicolumn{1}{l}{Weight Score} & \multicolumn{1}{l}{Normalized Weight} \\ \hline
\textbf{Cap} & Cap3 & 0.75 & 5 & 0.46 & Cap3 & 0.75 & 4 & 0.29 \\
\textbf{Cor} & Cor1 & 0.33 & 3 & 0.27 & Cor2 & 0.67 & 5 & 0.21 \\
\textbf{Com} & Com2 & 0.67 & 2 & 0.18 & Com2 & 0.67 & 4 & 0.29 \\
\textbf{Lc} & Lc3 & 1.0 & 1 & 0.09 & Lc2 & 0.67 & 3 & 0.21 \\ \hline
\textbf{$L_{DT}$} & \multicolumn{4}{c|}{0.6436} & \multicolumn{4}{c}{0.69} \\ \hline
\end{tabular}%
}
\end{table}

The corresponding maturity level for each dimension and assigned weight based on our understanding are presented in Table \ref{tab:Caseweight02}. The calculated dimensional maturity score and normalized weights are in the same table, which allows us to calculate the overall score for the two cases. As a result, the Tesla vehicle has a higher maturity level than Google Maps in the proposed framework, the difference mainly from the \textit{Capability} and \textit{Comprehensiveness} dimensions. However, the maturity level and weight score assignment for each dimension is primarily based on our understanding, while other practitioners may hold different opinions.

\begin{figure}
    \centering
    \includegraphics[width=0.5\linewidth]{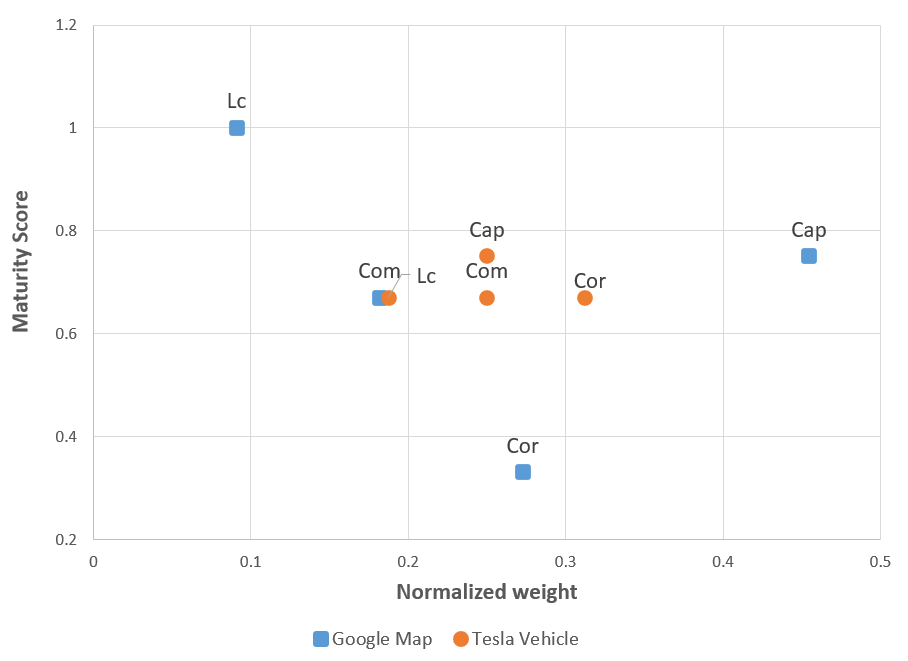}
    \caption{Result of evaluation for general cases, GoogleMaps and Tesla Vehicles}
    \label{fig:case22}
\end{figure}

Further evaluation can be visualized from Figure \ref{fig:case22}. In the case of Tesla vehicles, most dimensions have development levels proportional to their weights except from \textit{Life-cycle}. Thus, it may be worth further investment to avoid becoming the shortest stave in the Buckets Effect. Google Maps has a relatively lower score in all dimensions. One possible reason would be the unclear physical entity and its virtual counterpart in Google Map's representation of the real world. Plus, despite some similarities, Google does not claim that DT technology has been applied to Google Map Navigation.

\subsection{Scientific case studies}
Aside from the general cases, several papers are selected from the literature as a case study to validate the proposed framework. This selection aims to ensure both a temporal breadth and a diverse disciplinary representation, thus improving our study's representativeness.
Additionally, the selected papers are thoughtfully considered to align with the two fundamental conditions of the framework, \textit{comprehensiveness} and \textit{bidirectional connection}. The one-by-one validations will not be presented in this section to avoid redundancy.

Initially, DT was introduced in Prognostics and Health Management (PHM), then became popular in manufacturing but remained widely used in diagnosis and predictive maintenance. By 2021, the scope has widened to diverse applications, including management, logistics, smart cities, healthcare, and agriculture. The selection of papers respects the application and development of DT in these disciplines in the past few years.

\begin{table}
\centering
\caption{Summary of case study on the validation of the proposed evaluation framework, the Cap, Cor, Com and Lc are corresponding levels for the four dimensions defined in Section \ref{sec: Proposal}: $Cap$-Capability, $Cor$-Cooperability, $Com$-Comprehensiveness, $Lc$-Lifecycle}
\label{tab:caseStudy}
\resizebox{\textwidth}{!}{%
\begin{tabular}{ccl lcccc}
\hline
\textbf{Year} & \textbf{Index} & \multicolumn{1}{c}{\textbf{Domain}} & \multicolumn{1}{c}{\textbf{Desc}} & \textbf{Capability} & \textbf{Cooperability} & \textbf{Comprehensiveness} & \textbf{Lifecycle} \\ \hline
2018 & \cite{tao_digital_2018} & Maintenance & \begin{tabular}[c]{@{}l@{}}Prognostics and Health Management (PHM) \\ for wind turbine's gearbox\end{tabular} & Cap3 & Cor1 & Com1 & Lc1 \\
2019 & \cite{liu_novel_2019} & Health & DT health care for elder's personal health management & Cap4 & Cor1 & Com1 &  Lc1 \\
2019 & \cite{liu_digital_2019} & Manufacturing & \begin{tabular}[c]{@{}l@{}}sheet material Automated flow-shop manufacturing\\ system (AFMS): cutting, edge grinding, toughen etc.\end{tabular} & Cap2 & Cor2 & Com1 &  Lc1 \\
2020 & \cite{ivanov_predicting_2020} & Logistics & Global supply chain of a company selling lighting equipment & Cap2 & Cor2 & Com1 & Lc1 \\
2020 & \cite{lu_digital_2020} & Management & \begin{tabular}[c]{@{}l@{}}Asset management using DT of a centrifugal pumps \\ in the heating, ventilation and air-cooling\end{tabular} & Cap2 & Cor1 & Com2 & Lc1 \\
2020 & \cite{luo_hybrid_2020} & Maintenance & DT predictive maintenance for cutting tool life in CNC & Cap3 & Cor1 & Com2 & Lc1 \\
2020 & \cite{dembski_urban_2020} & Smart City & DT for traffic management in smart city & Cap3 & Cor2 & Com2 & Lc1 \\
2021 & \cite{aheleroff_digital_2021} & Maintenance & Maintenance   schedule of constructed wetlands & Cap3 & Cor1 & Com2 & Lc1 \\
2021 & \cite{liu_digital_2021} & Manufacturing & DT of hollow glass processing & Cap3 & Cor2 & Com2 & Lc2 \\
2019 & \cite{alves_digital_2019} & Agriculture & \begin{tabular}[c]{@{}l@{}}Sensing Change(monitoring) and \\ SWAMP (precision irrigation) Project\end{tabular} & Cap3 & Cor1 & Com2 & Lc2 \\
2021 & \cite{zhang_digital_2021} & Management & \begin{tabular}[c]{@{}l@{}}Predicting   the availability to \\ manufacture a hydraulic valve in job shop\end{tabular} & Cap3 & Cor2 & Com2 & Lc1 \\ \hline
\end{tabular}%
}
\end{table}

Despite the limited scope of the case studies, several insights can be drawn from the selected papers:
\begin{itemize}
    \item \textbf{Capability}: The \textit{Capability} level ranges from Cap2 to Cap3 in the case study, suggesting that most DTs are not limited to real-time only. Plus, there is a clear preference for Cap3 where DT is capable of making predictions. This might originate from DT's early focus on PHM, as well as its initially defined function \cite{grieves_digital_2014}. However, despite some visions for ``intelligent DT" \cite{grieves_intelligent_2022}, few DTs are able to handle decision-making, let alone make reasonable and explainable decisions.
    \item \textbf{Cooperability}: All DT use cases fall under the cooperability levels of Cor1 and Cor2, indicating a distribution without clear preference between basic and advanced cooperability levels. The final level of cooperability, Cor3 is not reached by any of these cases, which can be considered a perspective for ideal DT. On the other hand, there is no clear evolution concerning the distribution of maturity level on \textit{Cooperability}, which suggests that the choice of cooperability level primarily depends on the specific application needs in practice rather than a general trend.
    \item \textbf{Comprehensiveness}: Similar for \textit{Comprehensiveness}, all use cases belongs to either Com1 or Com2, leaving Com3 at present a vision of all-in-one DT for future. However, in the scope of the case study, the use of Com2 becomes more frequent in the latter years, especially in 2020 and 2021, hinting at an increasing preference for DTs with broader application coverage, thanks to general technological progress. 
    \item \textbf{Lifecycle}: The majority of the use cases are concentrated at Lc1, with only a few at Lc2. As the DT technology matures, more use cases might be entering a more advanced stage to cover more phases in their lifecycle.
\end{itemize}

To further validate the proposed framework, two papers \cite{lu_digital_2020}\cite{liu_digital_2021} are evaluated using the weight mechanism. The weight score is assigned to each dimension based on our understanding and assumption; the normalized weight is then calculated using the method presented in Section \ref{sec: Proposal} for later analysis. The results are shown in Table \ref{tab:weightCase2Sci}, with the dimensional maturity score and the calculated overall maturity score.

\begin{table}
\centering
\caption{Maturity calculation for two scientific cases, $Cap$-Capability, $Cor$-Cooperability, $Com$-Comprehensiveness, $Lc$-Lifecycle, $L_{DT}$-Overall maturity score}
\label{tab:weightCase2Sci}
\resizebox{\textwidth}{!}{%
\begin{tabular}{l|cccc|cccc}
\hline
\textbf{} & \multicolumn{4}{c|}{\cite{lu_digital_2020}} & \multicolumn{4}{c}{\cite{liu_digital_2021}} \\
 & \multicolumn{1}{l}{Maturity Level} & \multicolumn{1}{l}{Maturity Score} & \multicolumn{1}{l}{Weight Score} & \multicolumn{1}{l|}{Normalized Weight} & \multicolumn{1}{l}{Maturity Level} & \multicolumn{1}{l}{Maturity Score} & \multicolumn{1}{l}{Weight Score} & \multicolumn{1}{l}{Normalized Weight} \\ \hline
\textbf{Cap} & Cap2 & 0.5 & 5 & 0.33 & Cap3 & 0.75 & 4 & 0.4 \\
\textbf{Cor} & Cor1 & 0.33 & 2 & 0.13 & Cor2 & 0.67 & 2 & 0.2 \\
\textbf{Com} & Com2 & 0.67 & 4 & 0.27 & Com2 & 0.67 & 3 & 0.3 \\
\textbf{Lc} & Lc1 & 0.33 & 4 & 0.27 & Lc2 & 0.67 & 1 & 0.1 \\ \hline
\textbf{$L_{DT}$} & \multicolumn{4}{c|}{0.4779} & \multicolumn{4}{c}{0.702} \\ \hline
\end{tabular}%
}
\end{table}

% \begin{table}
% \centering
% \caption{Maturity calculation for two scientific cases, $Cap$-Capability, $Cor$-Cooperability, $Com$-Comprehensiveness, $Lc$-Lifecycle, $L_{DT}$-Overall maturity score}
% \label{tab:weightCase1}
% \resizebox{\textwidth}{!}{%
% \begin{tabular}{l|ccc|ccc}
% \hline
% \textbf{} & \multicolumn{3}{c|}{\cite{lu_digital_2020}} & \multicolumn{3}{c}{\cite{liu_digital_2021}} \\
%  & \multicolumn{1}{l}{\textbf{Maturity Score}} & \multicolumn{1}{l}{\textbf{Weight Score}} & \multicolumn{1}{l|}{\textbf{Normalized Weight}} & \multicolumn{1}{l}{\textbf{Maturity Score}} & \multicolumn{1}{l}{\textbf{Weight Score}} & \multicolumn{1}{l}{\textbf{Normalized Weight}} \\ \cline{2-7} 
% \textbf{Cap} & 0.4 & 5 & 0.33 & 0.6 & 4 & 0.40 \\
% \textbf{Cor} & 0.33 & 2 & 0.13 & 0.67 & 2 & 0.20 \\
% \textbf{Com} & 0.67 & 4 & 0.27 & 0.67 & 3 & 0.30 \\
% \textbf{Lc} & 0.33 & 4 & 0.27 & 0.67 & 1 & 0.10 \\ \hline
% \textbf{$L_{DT}$} & \multicolumn{3}{c|}{0.445} & \multicolumn{3}{c}{0.642} \\ \hline
% \end{tabular}%
% }
% \end{table}

\begin{figure}
    \centering
    \includegraphics[width=0.5\linewidth]{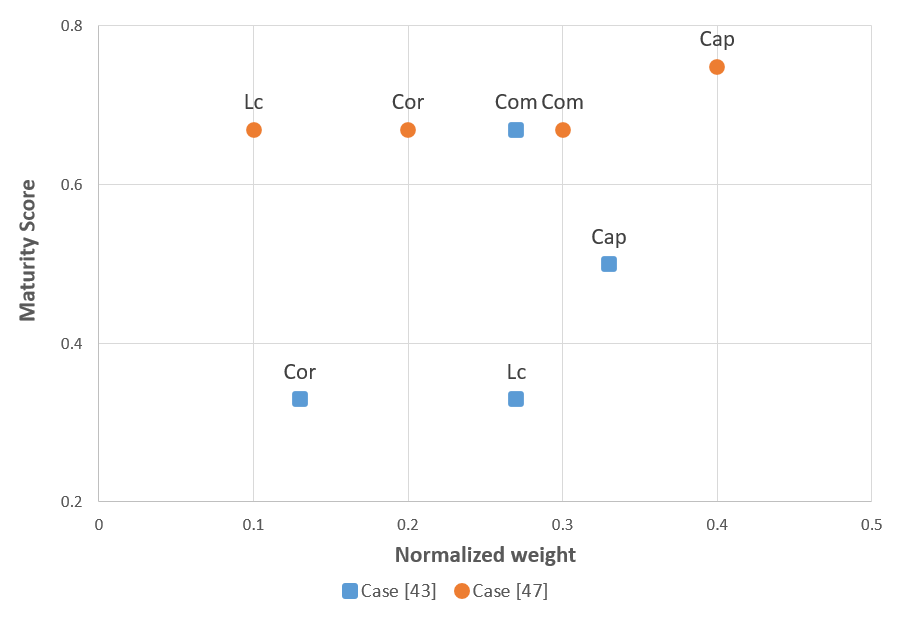}
    \caption{Result of evaluation in case \cite{lu_digital_2020}\cite{liu_digital_2021}}
    \label{fig:caseWS}
\end{figure}
Based on general maturity scores for DT cases are shown in Table \ref{tab:weightCase2Sci}, \cite{liu_digital_2021} is generally more developed than \cite{lu_digital_2020} based on the assigned weight.
The evaluation result for each dimension is shown in Figure \ref{fig:caseWS}, where the normalized weight and corresponding maturity score are presented for each dimension of the two cases. The dimension on the top left of the figure signifies low weight but high maturity, suggesting a possible need for resource reallocation, e.g., \textit{Lifecycle} dimension for case \cite{liu_digital_2021}. On the contrary, dimensions positioned in the bottom-right of the figure indicate high weight but exhibit low maturity. Therefore, they are likely to be the potential focus for future development, e.g., \textit{Lifecycle} and \textit{Cooperability} dimension for case \cite{lu_digital_2020}.

% \subsection{Beyond Scope}
~\\
As we are focusing on the evaluation of DT, there are, however, some related resources that exceed the scope of research, e.g., \cite{noauthor_iec_nodate}\cite{shao_analysis_2023}\cite{noauthor_system_nodate}. As shown in the citations, some works are from standardization organizations. They did not appear in the body of research as we focus on scientific articles in the literature review.

\section{Conclusion} \label{sec: conclusion}
The research question we are addressing is developing a unified DT evaluation framework to conform to practitioners' understanding of different disciplines. It paved the way for assessing general DT application from the modeling point of view, which is deemed the pivot of the evaluation strategy as models are considered the core of DT technology.

The motivation for this paper does not only come from the popularity of digital twin technology in academia and industry. In fact, despite its extending application in various domains over the years, few consensus has been reached on the definition and evaluation of DT, which would become a drawback for future work, which is the context of our research question. This paper makes two key contributions: firstly, a systematic literature review of existing DT evaluation strategies; and secondly, the introduction of a three-phase generic evaluation DT framework.

From the literature review based on the PRISMA method, we found that many calls for a globally accepted digital twin definition or standard; in the meantime, a lack of a commonly accepted evaluation tool to assess and design DT is revealed. Existing tools and models have diverse evaluation criteria choices, making them case-specific and hard to apply to other cases in different domains. 

Therefore, we have proposed a three-phase DT evaluation framework, aiming to achieve a certain level of consensus to build digital twins in different domains and guide future developers in deciding the appropriate DT for their application. The three corresponding phases are respectively: two fundamental conditions, four-dimensional classification schema, and calculation of the maturity score. 
Once validated by the two fundamental conditions, the use case pass to leveling phase in the four-dimensional classification schema. These dimensions are further adjusted by the weight mechanism based on their importance in case-specific requirements to ensure their adaptability. Qualitative and quantitative methods are combined to calculate the application case's dimensional and overall maturity scores. The evaluation model is then validated in general, industrial, and scientific case studies. The proposed model can help evaluate the existing DT use cases and be useful in aligning the understanding of stakeholders at the initial phase of DT development.

% \subsection{Need Enriching: Limitations and Perspectives}
\subsection{Limitations}
Currently, for the literature review of this article, only academic research from the Web of Science is utilized for analysis, suggesting that scientific articles from other databases might be missing. Meanwhile, industrial and commercial evaluation practices are not included in the literature. Future research may look into more databases, projects, and results from practitioners and businesses for a more representative and comprehensive finding. 

Additionally, the validation of the proposed evaluation framework is limited and hasn't been tested in real-world scenarios based on the practitioners' and experts' scoring and feedback. In the case studies, it is possible that the scores and weights we decided might fail to convince the related practitioners. The evaluation model itself, as well as the confirmation of maturity level and weight score, could be enhanced from expert knowledge in practical validation. It will become the focus of our future work to address this limitation and refine this framework through further engagement and collaboration with the people in relevant application domains.

\subsection{Perspectives}
As to conclude this research work, we must look into the future expansion of the proposed DT evaluation framework. The potential areas for further development in our perspectives are therefore listed as follows:
\begin{itemize}
    \item Future extensions of our framework may involve the integration of an additional technology maturity axis. This will allow us to capture and assess the evolution of the digital twin and its enabling technologies within the enhanced evaluation model.
    \item We also have plans to connect to the Capabilities Periodic Table from Digital Twin Consortium\footnote{https://www.digitaltwinconsortium.org/initiatives/capabilities-periodic-table/}, which provides a global view of the enabling technologies of DT throughout disciplines. Our framework could be improved by tool mapping, which concerns identifying and incorporating existing and emerging tools within the DT landscape. This could provide the practitioner with a comprehensive and interoperable toolbox in the design, assessment, and development practice. One of the possible approaches is a chatbot-style advisor, which can provide detailed and practical insights and recommendations based on the evaluation results to find the optimal combination solutions in the Capabilities Periodic Table. 
\end{itemize}
In summary, our proposed DT evaluation framework serves as a cornerstone to demystify the buzzword and move towards an aligned understanding of digital twins. The future of this work is revealed with opportunities to ensure the general applicability and adaptability in response to the rapid technological advancements and diverse application needs in the field of digital twins.

\section*{Acknowledgement}
This work is performed within the CHAIKMAT project funded by the French National Research Agency (ANR) under grant agreement “ANR-21-CE10-0004-01”.

% \footnotetext{https://www.digitaltwinconsortium.org/initiatives/capabilities-periodic-table/}

% If you have bib database file and want BibTeX to generate the
% bibitems, please use
%
\bibliographystyle{ieeetr} 
\bibliography{biblio02.bib}




% AI disclaimer
\section*{Declaration of generative AI and AI-assisted technologies in the writing process}
During the preparation of this work the authors used ChatGPT exclusively in order to provide suggestions to improve readability. After using this tool, the authors reviewed and edited the content as needed and take full responsibility for the content of the publication.

% \end{thebibliography}

\end{document}